\documentclass[openacc]{rstransa}

\usepackage[caption=false]{subfig}

\usepackage{bm}

\newcommand{\ie}{\emph{i.e.,~}}
\newcommand{\eg}{\emph{e.g.,~}}
\newcommand{\etal}{\emph{et al.~}}

\begin{document}



\title{Modern Perspectives on Near-Equilibrium Analysis of Turing Systems}

\author{
Andrew L. Krause$^{1,2}$, Eamonn A. Gaffney$^{1}$, Philip K. Maini$^{1}$ and V{\'a}clav Klika$^{3}$}

\address{$^{1}$Wolfson Centre for Mathematical Biology, Mathematical Institute, University of Oxford, Andrew Wiles Building, Radcliffe Observatory Quarter, Woodstock Road, Oxford, OX2 6GG, United Kingdom\\
$^2$Department of Mathematical Sciences, Durham University, Upper Mountjoy Campus, Stockton Rd, Durham DH1 3LE, United Kingdom\\
$^{3}$Department of Mathematics, FNSPE, Czech Technical University in Prague, Trojanova 13, 120 00 Praha, Czech Republic}

\subject{xxxxx, xxxxx, xxxx}


\corres{Andrew L. Krause\\
\email{andrew.krause@durham.ac.uk}}

\begin{abstract}
In the nearly seven decades since the publication of Alan Turing's work on morphogenesis, enormous progress has been made in understanding both the mathematical and biological aspects of his proposed reaction-diffusion theory. Some of these developments were nascent in Turing's paper, and others have been due to new insights from modern mathematical techniques, advances in numerical simulations, and extensive biological experiments. Despite such progress, there are still important gaps between theory and experiment, with many examples of biological patterning where the underlying mechanisms are still unclear. Here we review modern developments in the mathematical theory pioneered by Turing, showing how his approach has been generalized to a range of settings beyond the classical two-species reaction-diffusion framework, including evolving and complex manifolds, systems heterogeneous in space and time, and more general reaction-transport equations. While substantial progress has been made in understanding these more complicated models, there are many remaining challenges that we highlight throughout. We focus on the mathematical theory, and in particular linear stability analysis of `trivial' base states. We emphasise important open questions in developing this theory further, and discuss obstacles in using these techniques to understand biological reality. 
\end{abstract}


\begin{fmtext}

\end{fmtext}
\maketitle

\section{Introduction}
Alan Turing's \emph{Chemical Basis of Morphogenesis} \cite{turing1952chemical} has had profound impacts within mathematics, physics, chemistry and biology, especially in developmental and ecological settings \cite{ball2015forging}, as evidenced by the other articles in this Theme Issue. Nonetheless the approach taken in his 1952 paper used relatively straightforward mathematical techniques. Turing was aware of how simple of an approach he took to understanding the enormous complexity of biological development, stating, ``This model will be a simplification and an idealization, and consequently a falsification. It is to be hoped that the features retained for discussion are those of greatest importance in the present state of knowledge." We now have a far greater understanding of this biological complexity empirically, and have made progress in extending Turing's original models, and their analysis. This review will focus on these mathematical developments, concentrating on ideas which naturally generalize those present in Turing's work, as well as on open problems using these ideas to elucidate principles of self-organization in developmental biology and beyond.

While Turing's mathematical ideas about morphogenesis have been heavily extended, particularly including a much wider class of models and exploring them via numerical simulation, one of the simplest approaches to analyzing reaction-diffusion-like systems is the one employed by Turing: linear instability analysis of simple base states (which we refer to as near-equilibrium analysis) \cite{cross1993pattern, murray2004mathematical}. In this paper we will review developments of this aspect of the theory, suggesting how powerful such a classical approach can be when suitably extended, but also highlighting fundamental difficulties in studying pattern formation beyond these initial symmetry-breaking instabilities. We hope to point out fundamental obstacles in extending this kind of theory to more realistic scenarios of pattern formation, and particularly elucidate how such an approach could help us identify mechanisms and principles in biological pattern formation.

Beyond near-equilibrium analysis, an immense effort has been spent in studying far-from-equilibrium pattern formation in reaction-diffusion and related systems. While we will not review these approaches in detail, we mention two other articles in this Theme Issue which touch on some techniques for going beyond near-equilibrium analysis. In \cite{champneys}, the authors use a framework called \emph{spatial dynamics} to obtain a global picture of bifurcations in a wide class of systems, and suggest how various aspects of these bifurcation diagrams are in some sense universal. Spatial dynamics and related approaches study solutions of time-independent reaction-diffusion systems on $\mathbb{R}$ by thinking of the spatial variable as `time', and considering the two-component system as a flow in $\mathbb{R}^4$, where one can then make use of all of the machinery for such systems, such as numerical continuation \cite{yochelis2009towards, avitabile2017ducks, champneys2021bistability}.

Another important approach is the asymptotic study of localized solutions, sometimes referred to as a \emph{shadow-limit} of a reaction-diffusion system \cite{hale1989shadow,iron2001stability}. When one of the diffusion coefficients is very large, for certain classes of nonlinear reaction kinetics, one can find analytical spike solutions asymptotically which resemble highly-localized solitons. Linear stability analysis of such solutions involves the study of a nonlocal eigenvalue problem, which has been carried out analytically and numerically for a range of systems, geometries, boundary conditions and other extensions \cite{wei2013mathematical}. We also refer the interested reader to \cite{ward}, within this Theme Issue, for an overview of this approach in the context of bulk-surface systems.

In contrast to the approaches above, near-equilibrium analysis makes far fewer assumptions on the transport operator, geometry of the solution domain, and nonlinearities involved, and so in some sense is a more general kind of approach. There are important limitations to this technique, not least of which is due to how rich reaction-transport systems can be far from the equilibrium base state where these techniques are formally valid. Nevertheless, we hope to show how powerful this perspective can be in understanding a wide variety of problems in pattern formation for such systems. We will aim to give a broad review of Turing-type patterning in a range of contexts, but we will by no means be comprehensive. We will mostly focus on the near-equilibrium and related approaches to understanding these systems. Still, the choice of topics will surely still be idiosyncratic, and reflect the authors' personal tastes.

In Section \ref{linear} we provide an overview of the classical linear instability analysis for two-species reaction-diffusion systems, followed by a generalization of the ideas to the setting of $m$-component reaction-transport systems in a general state space. In Section \ref{structure} we show how these ideas have been applied to the study of pattern formation on manifolds and networks, as well as multi-domain models which require some further generalization of the basic ideas. In Section \ref{hetero}, we discuss reaction-diffusion systems with heterogeneity in space and time, and how linear instability theory can provide useful insights after suitable generalizations. Following this, in Section \ref{general} we discuss other kinds of reaction-transport systems that have been studied, such as those involving chemotaxis and cross-diffusion, as well as some important thermodynamic considerations. We also discuss both modelling and mathematical difficulties in going beyond two-species systems. Finally, we close in Section \ref{discussion} with a view towards the future, and in particular outline aspects of the problems presented which are ripe for further study. Throughout we hope to give a sense of how far we have come from Turing's initial insight, but also how much there is left to do in understanding these kinds of models, and their relationships to biological reality.


\section{Near-Equilibrium Analysis}\label{linear}
Here we review the classical Turing analysis of reaction-diffusion systems, essentially following the ideas in \cite{murray2004mathematical} which have become standard. We then show how, subject to some assumptions, these ideas readily generalize to reaction-transport systems. By reaction-transport system, we mean a generalization of reaction-diffusion systems with the Laplacian replaced by a general local transport operator, such as convection-diffusion, chemotaxis, or a graph Laplacian.

We first note that the restriction to systems of reaction-diffusion equations is important for the kind of pattern formation in which we are interested. While some scalar reaction-diffusion equations can admit non-constant equilibrium solutions, these do not arise as a direct interaction of nonlinear reactions and diffusion alone, but require other ingredients. Classic results show that scalar quasilinear reaction-diffusion equations with homogeneous Neumann boundary conditions in convex domains do not admit non-constant stable equilibria, even if temporally forced \cite{chafee1975asymptotic, matano1979asymptotic, hess1987spatial}. In fact, such non-existence of patterns has been shown for a large class of competitive and cooperative systems in convex domains \cite{kishimoto1985spatial}. Bistable equations in non-convex domains have been shown to exhibit stable non-constant solutions \cite{hale1984nonlinear}, as have bistable equations with Dirichlet or inhomogeneous boundary conditions \cite{murray1983minimum, murray2004mathematical}. Such solutions do not have the same qualitative features as patterned states in reaction-diffusion systems, or scalar equations with more general transport operators, which we focus on for the rest of this review.

\subsection{Linear Instability of Reaction-Diffusion Systems}\label{instab_RD}
The standard setting for a Turing instability is the two-component system,
\begin{equation}\label{first_RD}
    \frac{\partial u}{\partial t} = \nabla \cdot(D_u \nabla u) + f(u,v), \quad \frac{\partial v}{\partial t} = \nabla \cdot(D_v \nabla) v + g(u,v), \quad x \in \Omega, \,\, t>0
\end{equation}
where $u,v$ denote morphogen concentrations, $D_u,D_v>0$ diffusion coefficients, $f,g$ reaction kinetics, and the domain $\Omega \subset \mathbb{R}^n$ is usually taken to be a compact domain with sufficiently smooth boundary, such as a line for $n=1$ or a disc for $n=2$. We note that the `morphogens' in reaction-diffusion systems can be thought of as molecular species, such as the proteins Nodal and Lefty \cite{muller2012differential}, or can also correspond to other structures, such as cells in different morphological states, or other developmentally-important factors used for signalling \cite{kondo2010reaction}. This is typically augmented with no-flux or periodic boundary conditions on $u$ and $v$. The nonlinear kinetics are often polynomials or rational functions, as they typically arise from the application of Mass Action kinetics or other mean-field frameworks. A Turing instability of this system is said to occur when a homogeneous steady state (\eg$u^*, v^*$ such that $f(u^*,v^*)=g(u^*,v^*)=0$) is stable in the absence of the diffusion terms, but can become unstable when these are included. Writing $\mathbf{u} = (u,v)$ and $\mathbf{u^*} = (u^*,v^*)$, we linearize \eqref{first_RD} by letting $\mathbf{u} = \mathbf{u^*} +\epsilon \mathbf{U}$ for $|\epsilon| \ll 1$. Dropping higher-order terms we find,
\begin{equation}\label{linear_RD}
    \frac{\partial \mathbf{U}}{\partial t} = \mathbf{D} \nabla^2 \mathbf{U} + \mathbf{J}\mathbf{U},
\end{equation}
where $\mathbf{D} = \textrm{diag}(D_u,D_v)$, and $\mathbf{J}$ is the Jacobian of the functions $f$ and $g$ evaluated at the steady state $\mathbf{u^*}$.

Following \cite{murray2004mathematical}, we use an ansatz of the form,
\begin{equation} \label{Ansatz}
    \mathbf{U} = e^{\lambda_k t} w_k(\mathbf{x}) \mathbf{c_k},
\end{equation}
where $\mathbf{c_k} \in \mathbb{R}^2$, and $w_k$ are the eigenfunctions of the negative Laplacian with eigenvalue $\rho_k$ satisfying,
\begin{equation}\label{eigenvalue}
    \nabla^2 w_k + \rho_k w_k = 0.
\end{equation}
Throughout we will refer to $w_k$ as a mode, with corresponding growth rate $\lambda_k$. With homogeneous Neumann or periodic boundary conditions, we have that $\rho_0=0$ and $\rho_k$ can be arranged in a countable increasing sequence with $\rho_k \to \infty$ as $k \to \infty$. For many other kinds of boundary conditions (\eg Dirichlet, Robin), we can still obtain a discrete and increasing sequence of $\rho_k$, though we may have $\rho_0\neq 0$ leading to some subtleties when considering homogeneous base states and their stability \cite{klika2018domain}. However, in general homogeneous equilibria will not satisfy these more general kinds of boundary conditions, so despite a similar spectral problem, the analysis here would not apply to these cases. For concreteness, we only consider boundary conditions admitting spatially homogeneous equilibria (see Section \ref{hetero}\ref{spat_het} for approaches to linear instability analysis for spatially heterogeneous base states). 

This ansatz allows us to quickly analyze the stability of perturbations which evolve according to Equation \eqref{linear_RD}. With this substitution, a solvability condition (derived by requiring nontrivial values for $\mathbf{c_k}$) relates values of $\lambda_k$ to the eigenvalues $\rho_k$ and model parameters via,
\begin{equation} \label{disp_relation}
    \det(\lambda_k \mathbf{I} + \rho_k \mathbf{D} - \mathbf{J}) = 0,
\end{equation}
where $\mathbf{I}$ is the identity matrix. This determinant can be evaluated as a quadratic polynomial determining $\lambda_k$ for each distinct value of $\rho_k$. Of the two values of $\lambda_k$ for each $k$, we are typically interested in the one with largest real part.

For homogeneous Neumann or periodic boundary conditions, a Turing instability corresponds to the situation where $\Re(\lambda_0)<0$ and $\Re(\lambda_k)>0$ for some $k>0$. In this case, the two growth rates denoted by $\lambda_0$ correspond exactly to the eigenvalues of $\mathbf{J}$, and so the first requirement says that $\mathbf{J}$ is a stable (Hurwitz) matrix. More generally, analysis of the zeroes of the transport operator is needed to determine how to think about stability in the absence of diffusion \cite{klika2018domain}, but typically we require $\mathbf{J}$ to be a stable matrix. We consider the value of $\lambda_k$ as a function of the spatial eigenvalue $\rho_k$, and refer to this as the dispersion relation 

From this analysis, it can be shown that for two-species systems, the largest growth rate $\lambda_k$ will be real at the parameter space boundary of a Turing bifurcation (so that such a bifurcation does not indicate an oscillatory instability). Several other facts can readily be established about these instabilities, such as requiring $D_u \neq D_v$, $f_u$ and $g_v$ to be of opposite signs, with the positive sign corresponding to a slower-diffusing activator, and the negative a fast-diffusing inhibitor. This is a classical example of local-activation long-range-inhibition, which has been implicated in a range of biological pattern formation scenarios \cite{meinhardt2000pattern}. The Turing space can be defined from this analysis as the set of parameters which admit this kind of instability. For large domains (\ie in the approximation of a continuous spectrum $\rho_k$), one can show that such a space is defined by a set of four inequalities. We refer to  \cite[Chapter 2]{murray2004mathematical} for a systematic treatment of these classical results. 

  We remark that the standard ansatz in the stability analysis above is based on the existence of a complete orthonormal basis of the Laplace operator (including the boundary conditions) in $L^2(\Omega)$. This allows us to replace the study of the partial differential equation \eqref{linear_RD} by a study of an infinite system of coupled ordinary differential equations, and then exploit orthogonality to study these modes independently, justifying the ansatz \eqref{Ansatz}. There are important aspects to making this argument rigorous as decomposing the perturbation as a sum of eigenfunctions is taken in a $L^2$ limit sense, and hence the validity of exchanging the sum and temporal and spatial partial derivatives has to be shown. This can be done by considering a finite-dimensional Galerkin approximation, and showing suitable \emph{ a priori} estimates of the solution; we refer to \cite{evans1998partial} for further details. 
  

  The large time behaviour of the linear problem \eqref{linear_RD} is well captured by the above-mentioned system of ordinary differential equations, and hence the dispersion relation \eqref{disp_relation}, validating the use of the ansatz in the stability analysis. Alternatively, one can invoke properties of the heat semigroup and the spectral mapping theorem from the theory of semigroups\cite{engel2001one, pazy2012semigroups} to formalize this analysis. However, as the system is typically nonnormal, there are transient behaviours that may drive the system away from the region of validity of its linearisation, invalidating the analysis above. It was recently shown that in the classical two-species reaction-diffusion case, the significance of these transient effects is limited to the edge of Turing space and hence fine tuning of the parameters may be required to observe it \cite{klika2017significance}. In conclusion then, we can justify the ansatz \eqref{Ansatz}, and in particular make use of the dispersion relation \eqref{disp_relation} directly to assess the potential for a system to admit Turing instabilities.

  Finally, we briefly discuss what happens beyond the initiation of a Turing instability. In general the linear theory shows that specific eigenmodes (those with $\Re(\lambda_k)>0$) will exponentially grow from the base state, as long as they are excited by the initial perturbation to the spatially uniform steady state. Beyond this, the theory does not predict what happens asymptotically in time. In particular, while one can in principle take the $k$th eigenmode corresponding to the growth rate with largest real part, it is often the case that the band of unstable modes is large, especially when $D_v/D_u \gg 1$. In such cases, these modes will compete with one another, and one cannot say \emph{a priori} what wavelength the final pattern might have.
  
  Near the onset of the instability (\ie for parameters close to the Turing space boundaries), weakly nonlinear analysis can be used to study the growth, competition, and saturation of these modes \cite{cross1993pattern, wollkind1994weakly}. Such amplitude equations allow one to deduce whether the bifurcation is subcritical or supercritical. In the more well-studied supercritical case, these equations show the existence of small amplitude branches emerging from the homogeneous equilibrium, similar to a pitchfork bifurcation in a two-dimensional dynamical system. Generally these bifurcation branches persist far from parameter ranges wherein the analysis is formally valid, but they can undergo subsequent complicated bifurcations in general, limiting the applicability of the analysis away from the bifurcation boundaries. In the subcritical case, the linear analysis gives essentially no information on what kinds of patterns may emerge, as any spatially non-uniform equilibrium solution will not emerge from the homogeneous equilibrium, but exist some distance away from it. Such patterns can involve more exotic structures, such as pattern formation due to homoclinic snaking \cite{brena2014subcritical}, though the analysis of such patterns is beyond even weakly nonlinear theory. Still, the weakly nonlinear regime can be studied to deduce important qualitative features of the dynamics, such as determining what kinds of nonlinearities permit spot or stripe patterns in two spatial dimensions \cite{ermentrout1991}. We refer to \cite{wheeler2021convective} for a general application of these kinds of methods to a large class of equations, including reaction-transport systems.  See also \cite{champneys} in this Theme Issue for a discussion of generic features of these far-from-equilibrium bifurcations.

\subsection{Linear Instability of Reaction-Transport Systems}\label{RTS_sect}

We now demonstrate that the main ingredients of the above analysis can be applied to a more general reaction-transport system. We replace the spatial domain $\Omega$ either by a discrete set (such as a graph or lattice), or a submanifold of $\mathbb{R}^n$ for $n\geq 1$. We also replace the Laplacian by a scalar elliptic operator $\mathcal{L}$ which admits only a point spectrum, and assume that the corresponding eigenfunctions form a complete orthonormal basis of a suitable function space (typically $L^2(\Omega)$). Finally, we assume that $\mathcal{L}$ annihilates functions which are constant over $\Omega$.

We then consider the reaction-transport system for a vector-value function $\mathbf{u}(\mathbf{x},t) \in \mathbb{R}^m$,
\begin{equation}\label{RTS}
    \frac{\partial \mathbf{u}}{\partial t} = \mathbf{D} \mathcal{L} \mathbf{u} + \mathbf{f}(\mathbf{u}),
\end{equation}
where $\mathbf{D} \in \mathbb{R}^{2m} \times \mathbb{R}^{2m}$ is a constant positive-semidefinite transport matrix, and $\mathbf{f}$ are again the nonlinear reaction kinetics. The classical case studied in the previous subsection considers $m=2$, but for spatial operators $\mathcal{L}$ of order four and higher, scalar equations ($m=1$) are sufficient to observe pattern-forming instabilities (see Section \ref{general}\ref{beyond}). One can proceed as in the classical case above to study perturbations of a homogeneous base state $\mathbf{u^*}$. The linear reaction-diffusion problem \eqref{linear_RD} is replaced by an almost identical system with the Laplacian replaced by $\mathcal{L}$, and one can study its solutions via an ansatz like \eqref{Ansatz}, where now the spatial eigenfunctions satisfy,
\begin{equation}\label{eigenvalues_RTS}
    \mathcal{L} w_k + \rho_k w_k = 0.
\end{equation}
With the generalization of the Jacobian and identity matrices to account for $m$ components, we obtain a dispersion relation as in \eqref{disp_relation}, with the only difference in this case being the values of the spatial eigenvalues $\rho_k$, and in considering a polynomial of degree $m$ for each $\lambda_k$. 

While the analysis is nearly identical, there are important new phenomena which can arise in this general setting, and we will explore these throughout the examples in this paper. We also note that the boundary conditions for this system can be more complicated than classical periodic or homogeneous Neumann conditions, and will depend heavily on the physics of the transport being modelled. While chemotaxis and more general forms of nonlinear diffusion (or reaction-advection-diffusion) are not modelled by the system \eqref{RTS}, one can often deduce a similar dispersion relation to \eqref{disp_relation} by expanding solutions in terms of other orthonormal bases, \eg classical Laplace eigenfunctions, though the question of which basis functions to use becomes more delicate (see Section \ref{general}\ref{general_transport} for further discussion).  Finally, rigorously justifying the ansatz \eqref{Ansatz} can depend on $\mathcal{L}$ and $\Omega$, and we omit further discussion of this here.

\section{Structured Domains}\label{structure}

We now discuss reaction-transport systems on spatial domains $\Omega$ which are not simple subsets of Euclidean spaces. In particular, we focus on manifolds, networks, and multi-domain models. The first two can be thought of as examples of the generalized linear instability analysis given in Section \ref{linear}\ref{RTS_sect} (at least in the case of undirected networks), whereas the latter will involve less obvious generalizations of the classical theory.

\subsection{Curved Manifolds}

The study of reaction-transport systems on curved manifolds goes back to Turing's paper \cite{turing1952chemical}, where he employed spherical harmonics to discuss a hypothetical mechanism behind gastrulation. In fact, much of Turing's paper uses spherical symmetry as a major discussion point for why symmetry breaking instabilities must play a role in development. 

Hunding pursued this symmetry breaking question in the context of reaction-diffusion systems defined inside spheres and prolate ellipsoids \cite{hunding1980dissipative, hunding1983bifurcations}, though the curvature of the domain here was restricted to its boundary. Varea \etal studied a reaction-diffusion system on the surface of a sphere numerically, finding striking symmetries in the emergent patterns \cite{varea1999turing}.  Chaplain \etal developed a numerical method to study a reaction-diffusion system on the surface of the sphere \cite{chaplain2001spatio} using an expansion of the solution in terms of spherical harmonics. They postulated that such pattern-forming systems may play a role in determining how tumour spheroids grow and develop. 

To model diffusion on surfaces and curved domains beyond simple curvilinear examples, one can introduce some machinery from differential geometry. The classical model of Fickian diffusion can be extended to a general compact orientable Riemannian manifold (for instance, a smooth two-dimensional surface) by replacing the Laplacian with the Laplace-Beltrami operator, which can be written in coordinates on the manifold in terms of a metric tensor. This exactly recovers the Laplacian on the surface of spheres, cylinders, and other curvilinear coordinate systems, but allows for models to be posed on more general surfaces and domains, particularly ones where a single global coordinate system may not be possible. The book by Frankel \cite{frankel2011geometry}, for example,  provides a general introduction to calculus on manifolds, including transport equations posed on them.

Plaza \etal studied reaction-diffusion systems on a variety of one- and two-dimensional manifolds \cite{plaza2004effect}, and discussed some general aspects of these models (including the role of growth in such systems, which we defer to Section \ref{structure}\ref{growth_sect}). They noted that the Laplace-Beltrami operator on such a manifold has essentially the same properties as the Euclidean Laplacian with periodic or Neumann boundary conditions, and hence deduced exactly the dispersion relation \eqref{disp_relation}. A further generalization involves including tangential divergence operators representing advection alongside diffusion modelled by the Laplace-Beltrami operator. Krause \etal studied reaction-advection-diffusion systems on the surface of a sphere, involving both Laplace-Beltrami and tangential divergence operators \cite{krause_emergent_2017}, finding that advection on such a closed manifold could give rise to spatiotemporal oscillations of pattern creation and destruction. Another recent application of these kinds of systems is the study of reaction-advection-diffusion models of vegetation and rainfall on curved terrains \cite{gandhi2018topographic, tzou2020analysis}.

\begin{figure} \centering
    \subfloat[$L=1$] {\includegraphics[width=0.3\textwidth]{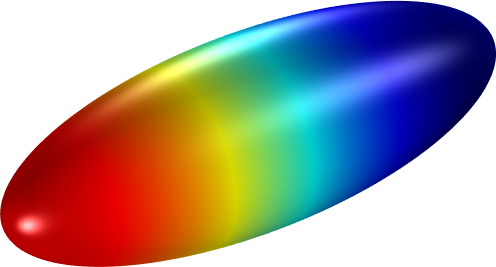}} \hspace{0.5cm}
    \subfloat[$L=2$] {\includegraphics[width=0.3\textwidth]{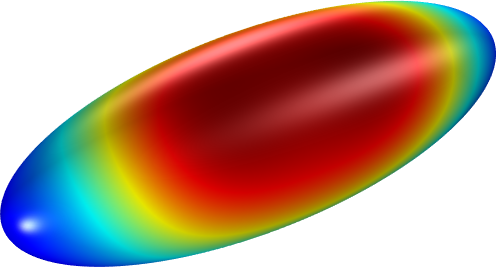}} \hspace{0.5cm}
    \subfloat[$L=20$] {\includegraphics[width=0.3\textwidth]{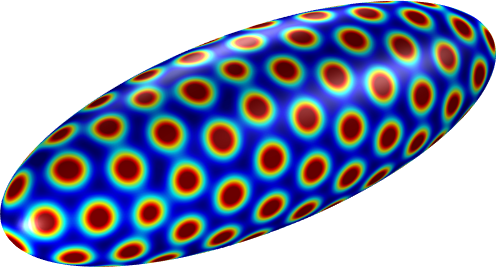}} 
  \caption{Plots of the activator $u$ from simulations of the Schnakenberg reaction-diffusion system on the surface of a prolate ellipsoid of semi-minor radii $L$, and semi-major radius $3L$. The kinetics are given by $f = 0.01-u+u^2v$, $g=1.7-u^2v$, with diffusion coefficients of $D_u=1$ and $D_v=30$. Simulations were carried out in Comsol using 20,460 triangular boundary elements. Initial data were taken as the homogeneous base state $u^*=1.71$, $v^*=0.58138$ with normally distributed noise with variance $10^{-1}$ added at each finite element, and these were evolved until $t=10^4$. The colour scale is fixed such that the dark red region is a maximum with $u =3.5$, and the dark blue region is a minimum with $u=0.5$.} 
  \label{Fig1}
\end{figure}

While the basic instability analysis is unchanged for Laplace-Beltrami operators compared with the standard Laplacian, curvature can influence mode selection as well as the structure of patterned states. As the relative domain size increases (or, as diffusion parameters are decreased), the structure of observed patterns becomes similar to those on large Euclidean domains, due to the relative size of curvature compared to the scale of the localized patterns. See Figure \ref{Fig1} for an example on a prolate ellipsoid. For $L=1$ and $2$, we see a single localized spot form which deforms around the manifold, whereas for $L=20$, many smaller spots form across the surface with little apparent sensitivity due to the curvature. Such finite-size effects of curvature may play important roles in cellular and developmental settings where only a few localized elements emerge, such as in cell polarization. However, beyond predicting when we might expect to see such finite size effects, linear analysis does not appear to have been successfully exploited to date in describing how localised patterns are deformed by curvature. Single-mode patterns may play crucial roles in certain models of cell polarization, where many results have already implicated geometry as important in these processes; see Section \ref{structure}\ref{multidomain}. More thoroughly exploring such situations is one important direction for future work.

While the linear stability analysis of homogeneous equilibria for reaction-diffusion systems on curved manifolds is a mild extension of the analysis on one-dimensional domains, many of the tools for analyzing pattern formation away from the homogeneous steady state do not have straightforward extensions to manifolds. As an example, a particular spatial eigenvalue on the sphere will, in general, have many distinct eigenfunctions, often referred to as degenerate due to their multiplicity. This makes predicting pattern structure from the fastest growing mode impossible, and complicates the analysis used to understand mode competition and pattern stability. See \cite{callahan2004turing, callahan2004turing, trinh2016dynamics} for details and examples of how this degeneracy in the eigenspace of the Laplace-Beltrami operator complicates such analysis. Beyond degeneracy issues, there are many open questions about the influence of curvature and manifold geometry more generally on the emergence and stability of far-from-equilibrium patterns, especially in the case of structures which are not highly localized, such as stripes.

\subsection{Reaction-Diffusion Systems on Networks}

Turing's paper also analyzed the case of a ring of discrete cells coupled by diffusion, with reactions confined to the nodes. Recent chemical experiments have suggested that such a system of discrete cells can give rise to the patterns predicted \cite{tompkins2014testing}. Othmer and Scriven extended Turing's ideas about discrete cells to reaction-diffusion systems on general graphs, with a particular emphasis on lattices and regular graphs in \cite{othmer1971instability}. Such discrete lattice models have also been used to conceptualize non-diffusible morphogens within cells, in comparison to those which can diffuse elsewhere \cite{rauch2004role}.  More recently, the authors in \cite{nakao2010turing, wolfrum2012turing, mccullen2016pattern, ide2016turing} and others have considered these discrete reaction-diffusion systems on a variety of complex network architectures, and used several numerical and analytical techniques to determine the role of topology in the emergence and structure of inhomogeneous steady states analogous to localized patterns in the continuous case. 

Given an undirected weighted graph (that is, with $n$ nodes or vertices $\bf{v}$ and $\ell$ edges $\bf{e}$), we can use the adjacency matrix $A_{ij}$ to describe a diffusion process between these nodes. We consider $m$ species at each node $v_i$ given by $\mathbf{u}_i$. At each node $i$ the species interact locally via the kinetics $\mathbf{f}(\mathbf{u}_i)$ and diffuse to neighboring nodes. The reaction-transport system \eqref{RTS} on a graph then looks like,
\begin{equation}
\frac{\textrm{d} \mathbf{u}_i}{\textrm{d} t} = \sum_{j=1}^{n}A_{ij}\mathbf{D}\left(\mathbf{u}_j-\mathbf{u}_i\right)+  \mathbf{f}(\mathbf{u}_i) = \sum_{j=1}^{n}L_{ij}\mathbf{D}\mathbf{u}_j+  \mathbf{f}(\mathbf{u}_i),
\end{equation}
where $\bf{L}$ is the graph Laplacian. As this is a system of $nm$ equations, one could directly construct a Jacobian from the linearization about an equilibrium of the kinetics and compute the eigenvalues. Alternatively, eigenvectors of the graph Laplacian (given by \eqref{eigenvalues_RTS}) can be used to directly deduce the same dispersion relation given by \eqref{disp_relation} in the analysis of a homogeneous steady state (\ie where $\mathbf{u}_i^*=\mathbf{u}_j^*$ for all $i,j=1\dots,n$). 

In this case there will be finitely many eigenvalues and eigenvectors, and so in some sense the analysis is simpler than in the spatially continuous case. In general, however, networks can have much more irregular topologies than spatially continuous domains, and understanding what features of the topology play a role in diffusion-driven instabilities is a major theme of current research \cite{asllani2014turing, kouvaris2015pattern, asllani2016tune}. A common approach in these analyses is to study how structure in the network topology influences the spectrum of $L$, which can then immediately impact whether or not a system admits a Turing instability via the dispersion relation \eqref{disp_relation}.  Multistability of the reaction kinetics can also play a role in forming patterns, as there is no longer a simple criterion, such as convexity, which prevents such states from coexisting in the presence of diffusion \cite{kouvaris2016self}.

\begin{figure}
    \centering
    \includegraphics[width=0.7\textwidth]{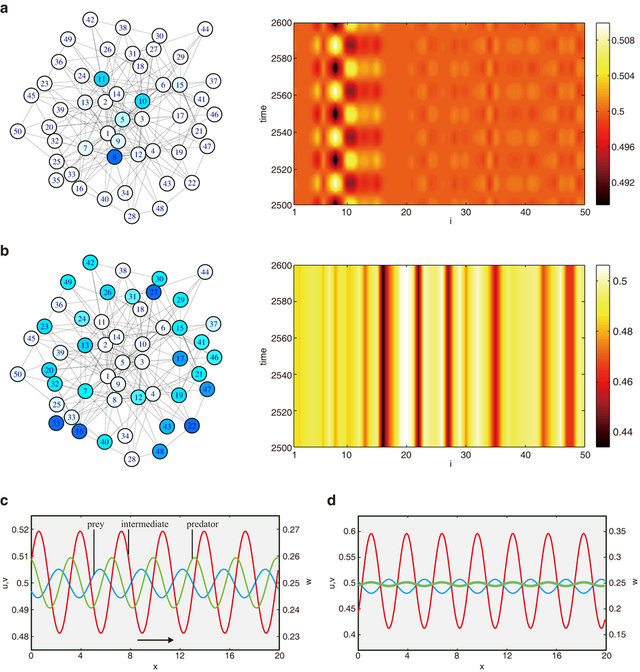}
    \caption{Examples of temporally oscillating solutions (a,c) and stationary Turing patterns (b,d) in networks (a,b) and in the respective continuous media with periodic boundary conditions (c,d), corresponding to numerical solutions of a networked predator-prey model. In panels (a) and (b), mean oscillation amplitudes for all network nodes are displayed on the left, and time evolution diagrams on the right. For continuous media, instantaneous concentration profiles are shown in (c,d) on a periodic domain. Reproduced with no changes from \cite{hata2014dispersal} under a \href{https://creativecommons.org/licenses/by/3.0/}{CC BY 3.0 License}. }
    \label{FigNetworks}
\end{figure}

Another difficulty in the setting of networks is that of visualizing Turing patterns. One example of an illuminating visualization is that found in Figure 2 of \cite{hata2014dispersal}, which we have reproduced in Figure \ref{FigNetworks}. We note that the usual periodicity found in Turing patterns (\eg as in Figure \ref{Fig1}(c)) is absent, as the network lacks the same isotropy as the spatially continuous models. We remark that the oscillatory solutions in Figure \ref{FigNetworks}(b,d) arise due to complex conjugate pairs of eigenvalues with $\Re(\lambda_k)>0$, and we will discuss such dynamics in more detail in Section \ref{general}\ref{beyond}.

An important point is that for an undirected graph, the graph Laplacian $\mathbf{L}$ will be real and symmetric, and hence normal. Directed networks, on the other hand, give rise to a much more complicated analysis. In general, the spatial eigenvalues $\rho_k$ will no longer be purely real, and existence of a basis of eigenvectors is no longer guaranteed. Nevertheless, such systems are being explored as a plausible mechanism for pattern formation in discrete networks that may have fewer of the restrictions (\eg differential diffusion) compared to the classical case \cite{muolo2019patterns}.

\subsection{Multi-Domain Models}\label{multidomain}

There have been recent efforts to study pattern formation in domains with multiple distinct regions. Examples of such systems include proteins which can both diffuse within a cell and bind to a membrane \cite{ward}, bacterial pattern formation on inert substrates \cite{grant2016orthogonal, krause2020turing}, and epithelial-mesenchymal signalling \cite{cruywagen1992tissue, shaw1990analysis}. 

 Broadly, we can consider multi-domain systems as being modelled in three distinct ways: instantaneously coupled models, bulk-surface models and stratified (or layered) models. The first are models where the components are assumed to occupy the same spatial domain (or the reactions occur in thin regions where a homogenisation approximation is appropriate) \cite{yang2002spatial, epstein2007coupled, yang2003oscillatory, fujita2013pattern}. Such models are essentially reaction-diffusion systems that have more components with linear coupling between subsystems, and are amenable to block-matrix analysis in the study of Turing instabilities  \cite{catlla2012instabilities}. A second class of model considers bulk-surface coupling, where the surface is sufficiently thin such that transverse (normal to the boundary) concentration gradients can be neglected in the surface model, but transverse gradients in the bulk can be significant. Such settings have been explored extensively, and are particularly suitable for models of proteins diffusing in the cytoplasm and binding on the cell membrane \cite{ratz2014symmetry, madzvamuse2015stability, spill2016effects, cusseddu2018coupled, paquin2018pattern, halatek2018review, frey2018review, paquin2020pattern}. These models have a variety of applications, but of particular note we mention cell polarization in mitosis and other contexts, where a single mode becoming unstable is sufficient to break symmetry \cite{wu2016geometry, brauns2020bulk}. We refer especially to \cite{ward} in this Theme Issue, as a review of these systems.
 
 Turing-type analysis has been considered in bulk-surface systems, primarily in the case of disc and spherical geometries \cite{levine2005membrane, madzvamuse2015stability}. In these cases, subject to a compatibility condition required for the existence of a spatially homogeneous steady state, one can make use of polar coordinates and an ansatz analogous to \eqref{Ansatz} with radial and angular coordinates separated in order to study the stability of homogeneous equilibria. One obtains a solvability condition similar in spirit to the dispersion relation \eqref{disp_relation}, though with additional complexity due to the coupling. The key feature which allows for such an approach is that the problem is entirely separable into radial and angular coordinates using the usual Laplace eigenfunctions. Beyond spherical geometries, we are unaware of the use of analytical techniques to extend this basic ansatz, though in principle any geometry amenable to a separable polar coordinate system (\eg parabolic or elliptical polar coordinates) would allow for a similar approach.
 
\begin{figure} \centering
    \subfloat[Bulk-Surface Model] {\includegraphics[width=0.45\textwidth]{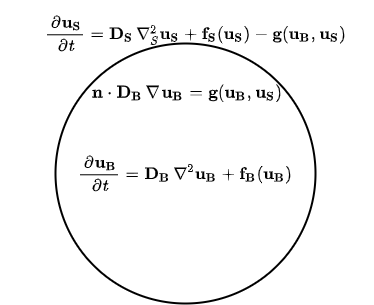}} \hspace{0.5cm}
    \subfloat[Stratified Model] {\includegraphics[width=0.45\textwidth]{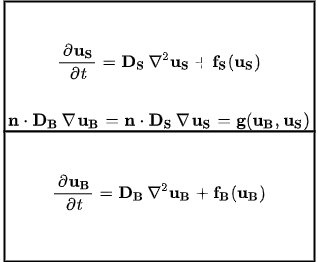}}
  \caption{Diagrams of bulk-surface (a) and stratified (b) models. Each system involves a bulk and surface species, $\mathbf{u_B}$ and $\mathbf{u_S}$ respectively, defined in distinct regions with their own diffusion coefficients and kinetics. The species' fluxes are coupled along a codimension-1 interface (a curve in these 2D examples) via a function of both concentrations along the interface, $\mathbf{g}$. In the bulk-surface case, the kinetics of the surface species are also modified by $\mathbf{g}$. This also explains the notation $\nabla^2_S$ which is the Laplacian restricted to the surface.} 
  \label{FigBulkSurface}
\end{figure}

 Finally, we consider models of stratified media, with two or more separated spatial domains with interfaces and suitable coupling boundary conditions. See Figure \ref{FigBulkSurface} for a diagram comparing these two modelling frameworks. Such models can sometimes be reduced to bulk-surface models given appropriate distinguished limits and scaling assumptions (for example, in \cite{fussell_hybdrid_2018} the authors reduce 2D spatially continuous regions to 1D boundaries assuming thinness of these regions).  Krause \etal \cite{krause2020turing} analyse a model of a two-layered stratified system in two spatial dimensions, primarily focusing on the case of a standard two-component system in the surface coupled to an inert layer which only allows for diffusion. This was motivated by an experimental framework of bacteria placed on top of agar (for use in synthetic pattern formation), and the study investigated the impact of this inert bulk on the pattern-forming potential of the surface reaction-diffusion system. Unlike in the analysis described above for radially-symmetric bulk-surface systems, one can not expand perturbations in terms of classical eigenfunctions, as these will not in general satisfy the coupling condition. Rather, an ansatz of the form \eqref{Ansatz} is used, but with an \emph{unknown} eigenfunction in the direction normal to the coupling interface. Solutions for these eigenfunctions are then found by treating the growth rates $\lambda_k$ as a parameter. This analysis leads to a dispersion relation involving transcendental functions of matrices, which can then be analyzed numerically and in terms of various asymptotic limits. While the dispersion relation successfully predicted pattern forming instabilities in the full system, it was far more difficult to understand (in terms of how terms in the model influenced pattern formation) than the classical polynomial dispersion relation \eqref{disp_relation}, and even challenging to solve numerically for the growth rates $\lambda_k$.
 
 This obstruction to the classical analysis (\ie eigenfunction expansions leading to polynomials analogous to \eqref{disp_relation}), arises from inseparability of spatial eigenfunctions due to the coupling. We conjecture that such an obstruction is generic for stratified models, even in other geometries. In particular, if we consider the circular and spherical geometries analyzed in \cite{madzvamuse2015stability}, but extend the surface outwards (so that we have an annular region coupled to an interior spherical one), we can no longer use the power series ansatz involving classical eigenfunctions, as the radial and angular eigenfunctions do not decouple. In principle, the more elaborate approach used in \cite{krause2020turing} could be employed, though it would be cumbersome in terms of calculations. As far as we are aware,  most multi-domain systems on generic geometries, both bulk-surface and stratified (bulk-bulk), will not admit analysis by classical series expansions of eigenfunctions, at least in a way that leads to polynomial dispersion relations. In other words, bulk-surface systems with boundaries conforming to polar coordinates seem to be a special class of models where the classical analysis generalizes in a straightforward way.  Such a situation is analogous to the coordinate-separability problems in analytically finding Laplace eigenvalues on polygonal domains, for which rectangular, elliptical, and equilateral triangular domains are the most prominent examples where simple analytical approaches suffice. Beyond these examples, almost all polygonal domains (including the hexagon) require numerical study \cite{colbrook2018computing}. 
 
 While some biological systems, such as some types of cells, can be idealized as bulk-surface systems on spheres (or more generally ellipsoids), we note that such geometries are often idealizations. Hence, we view these geometric complications to studying bulk-surface coupling to be of interest for investigating more realistic biological systems, such as in the case of cells which undergo substantial shape changes during migration or mitosis. Mathematically, these fundamental geometric challenges are interesting, especially in contrast to the generality of linear instability analysis in other settings (\eg curved manifolds) where the geometry can entirely be encoded in the spatial spectrum of the transport operator, $\rho_k$. Of course, there may be ways forward in the analysis of such problems that are simply not apparent from our limited point of view.
 
 There are other important problems in the case of multi-domain reaction-diffusion models. In the most general setting of reactions occurring in coupled regions, one typically obtains a set of consistency conditions for there to exist spatially homogeneous steady states, and in general such steady states will not exist. Even in the linear analysis of bulk surface models, one requires compatibility conditions to guarantee the existence of a homogeneous equilibrium \cite{madzvamuse2015stability}. As another example, if the inert region in the model described above in \cite{krause2020turing} is replaced by a region with linear degradation (likely a more physical assumption), then these compatibility conditions cannot be satisfied, so no positive homogeneous steady states would exist. Nevertheless, such a situation with reactions in multiple regions is likely the most common in biological applications, for both membrane-bulk and bulk-bulk systems. We remark that this is one situation where the far-from-equilibrium approaches, particularly those involving asymptotic analysis of the stability of spike solutions, is capable of handling these aspects better than the near equilibrium approach. See \cite{ward} and the references therein for examples where substantial progress can be made using these approaches. Finally we remark that in some asymptotic limits, these multi-domain models can be viewed as spatially heterogeneous systems where the kinetics or diffusive fluxes change throughout the domain. We will discuss such spatially heterogeneous models in the next Section.

\section{Heterogeneity}\label{hetero}

Turing instabilities are often understood as a route towards symmetry-breaking of homogeneous states, and pattern formation is often viewed as the emergence of spatial structure from a uniform background or initial state. In recent years, however, the basic ideas have been extended to reaction-diffusion systems in heterogeneous media, and to systems which are forced in time. In addition to what we discuss below, we also refer to  \cite[Chapter 11]{mendez2010reaction} for an overview of the literature on such systems. From the point of view of near-equilibrium analysis, both of these scenarios require a novel definition of a base state in order to define what one means by an emergent pattern. Throughout, we will mostly review work which studies models of the form \eqref{first_RD} where the diffusion coefficients and/or the reaction kinetics explicitly depend on space $\mathbf{x}$ and/or time $t$. Heterogeneity can also appear due to inhomogeneous or more complex (\eg Dirichlet or Robin) boundary conditions \cite{dillon1994pattern, maini1997boundary, tzou2011stationary,krause2020isolating}.

While many of the techniques for the analysis of such systems are recent developments, Turing himself was aware that such models would be more realistic in many settings, saying \cite{turing1952chemical} ``Most of an organism, most of the time, is developing from one pattern into another, rather than from homogeneity into a pattern. One would like to be able to follow this more general process mathematically also."  We also remark that domain growth, as well as spatial and temporal forcing terms, were crucial in Turing's later work on Fibonnaci phyllotaxis which he was unable to complete before his death; see \cite{swinton2004watching} for a thorough discussion. As these notes were largely ignored, and computation is invaluable in studying such systems, it was some decades before researchers began looking at the impact of growth and spatial heterogeneity in reaction-diffusion patterns.

\subsection{Spatially Heterogeneous Domains}\label{spat_het}

\begin{figure}
    \centering
    \includegraphics[width=0.9\textwidth]{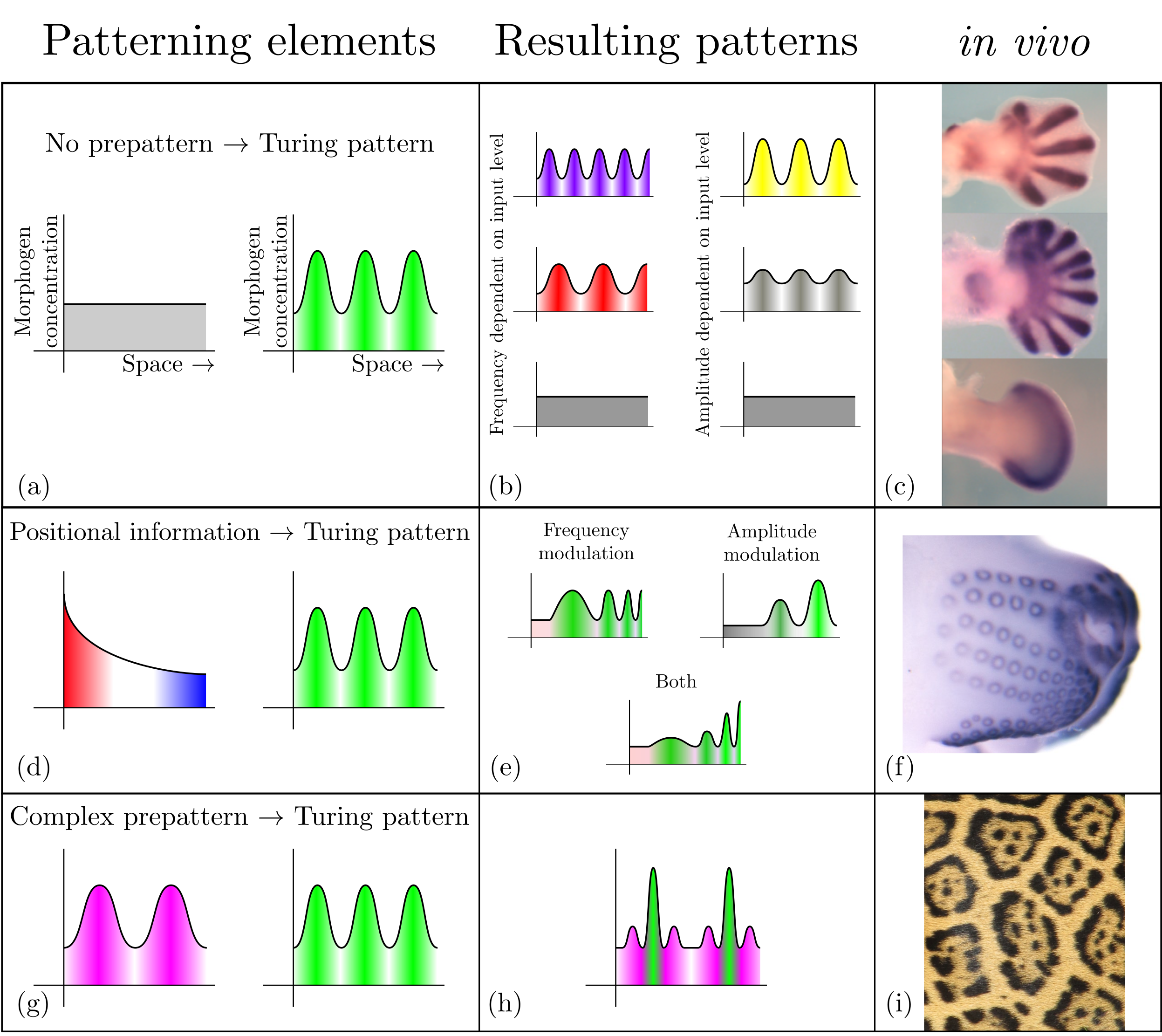}
    \caption{Different interactions of pattern formation mechanisms in development. (a) is a generic schematic of Turing pattern formation from homogeneity, with different pattern characteristics shown in (b), and, in (c), a biological example of a developing mouse paw in the presence of altered levels of Hox gene action. Positional information feeding into reaction-diffusion is shown in (d)-(e), consistent with observed structural characteristics of mouse whisker placodes in (f). Finally, successive reaction-diffusion patterning is shown in (g)-(h), with the example of Jaguar spots demonstrating large and small-scale pattern formation in (i).    In particular,  the schematic in (g) shows a sinusoidal prepattern (left peaks) feeding into a wave mode 3 Turing pattern (right peaks)   with, here for illustrative purposes,  the Turing pattern only able to   form  within the peaks of the prepattern. Thus, each peak forms a disjoint interval. Mouse paw images from \cite{Sheth1476} R. Sheth, L. Marcon, M. F. Bastida, M. Junco, L. Quintana, R. Dahn, M. Kmita, J. Sharpe, M. A. Ros. Hox genes regulate digit patterning by controlling the wavelength of a turing-type mechanism. Science, 338(6113):1476–1480, 2012. Reprinted with permission from AAAS. Mouse whisker placode image used with permission from Denis Headon. Jaguar picture by Jean Beaufort used under a CC0 Public Domain license from http://bit.ly/JaguarPicture. Figure reproduced from \cite{krause_WKB}; used with permission.}
    \label{FigHet}
\end{figure}

Beyond Turing's work on phyllotaxis, spatial heterogeneity was also used by Gierer and Meinhardt in their classical work on pattern formation  \cite{gierer1972theory}. Heterogeneity in developmental settings has been suggested as key for organising different regions along cell boundaries based on sharp variations in gene expression \cite{meinhardt1983cell,meinhardt1983boundary,irvine2001boundaries,krause2020isolating}. More recent work has used spatial heterogeneity in reaction-diffusion models to relate Turing-type pattern formation to other patterning theories, such as positional information \cite{green2015positional, wolpert2016positional, krause_WKB}. A complete theory of multi-scale patterns is still to be developed, but these advances help provide plausible explanations for biological patterns with periodic patterning at multiple scales; see Figure \ref{FigHet} for an overview.

Besides pattern formation in development, models of spatially heterogeneous reaction-diffusion systems have been employed to study collective animal dispersal \cite{pickett1995landscape, clobert2009informed, cobbold2015diffusion, bassett2017continuous, kurowski2017two}, reaction-diffusion in domains with non-isotropic growth \cite{crampin2002pattern, krause2019influence}, as well as spatial invasion modelling \cite{sun2016pattern,belmonte2013modelling}, and models with differential diffusion leading to spatial inhomogeneity in plant root initiation \cite{brena2015stripe, avitabile2018spot}, among other applications. Classical mathematical work on heterogeneous reaction-diffusion systems can be found in \cite{auchmuty1975bifurcation}, where they study bifurcations of solution branches as parameters vary. Numerical and asymptotic studies have shown that heterogeneity can change local instability conditions for pattern formation \cite{benson1993diffusion, page2003pattern}, modulate size and wavelength of patterns \cite{page2005complex}, and localize (or pin) spike patterns in space \cite{iron2001spike, ward2002dynamics, wei2017stable}. There is also a large literature on reaction-diffusion systems with strongly localized heterogeneities \cite{yuan2007heterogeneity, doelman2018pulse}, and numerous studies exploring spatially heterogeneous reaction-diffusion systems in chemical settings \cite{lengyel1991modeling,epstein1996nonlinear,rudiger2003dynamics,miguez2005turing,rudiger2007theory, yang2002spatial,peter2005stripe,haim2015non}. See \cite{epstein} in this Theme Issue for a modern review of chemical approaches to studying Turing systems.

There has been some work to develop a theory generalizing the classical Turing instability analysis in Section \ref{linear}\ref{instab_RD} to the case of spatially heterogeneous systems. There are challenges in extending the classical analysis, both because of the difficulty of defining and solving a suitable spectral problem, and of determining the most relevant `base' state to perturb. Several papers have used particular tools to analyze heterogeneous solutions exploiting special asymptotic assumptions and specific nonlinear kinetics \cite{dewel1989effects, kuske1997pattern, otsuji2007mass}, as well as spatially varying diffusion coefficients of a particular form \cite[Chapter 11]{mendez2010reaction}.  One approach to obtaining general results is the idea of using Galerkin expansions of standard functions, as discussed in Section \ref{linear}\ref{instab_RD}, and then studying a suitably truncated system (where modes are no longer decoupled due to the heterogeneity). This idea was used by Dillon \etal \cite{dillon1994pattern} to aid a numerical continuation study of inhomogeneous and mixed boundary conditions. More recently, Kozak \etal \cite{kozak2019pattern} employed this approach to study systems with piecewise constant reaction kinetics, finding that one can broadly use the intuitive homogeneous Turing conditions away from the discontinuity in the domain (with a base state that itself is piecewise homogeneous). Finally we remark that \cite{vangorder}, in this Theme Issue, provides a significant generalisation of this Galerkin approach to a range of systems with spatial heterogeneity in kinetics, diffusion coefficients, and boundary conditions.

In a similar direction to \cite{kozak2019pattern}, Krause \etal \cite{krause_WKB} deduced a set of `local' Turing conditions with arbitrary spatial heterogeneity in the kinetics, under the assumption that the heterogeneity is sufficiently smooth and not rapidly varying compared with the diffusive length scales. The reaction-diffusion system for many models can then be scaled to be of the form,
\begin{equation}
    \frac{\partial \mathbf{u}}{\partial t} = \varepsilon^2\mathbf{D}\frac{\partial^2 \mathbf{u}}{\partial x^2} + \mathbf{f}(x,\mathbf{u}),
\end{equation}
where $\varepsilon \ll 1$, $x \in [0,1]$, and all other parameters are $O(1)$ with respect to $\varepsilon$. For such a system, a natural base state is the heterogeneous solution to $\mathbf{f}(x,\mathbf{u^*})=0$. One can then analyze the stability of such a base state using a WKBJ ansatz for the perturbations, and show that in the limit of small $\varepsilon$, the stability of the base state $\mathbf{u^*}$ is entirely determined by whether or not the classical Turing conditions are satisfied locally. While these localisation results are consistent with the results in \cite{kozak2019pattern} for discontinuous (piecewise constant) heterogeneity, we remark that to justify either of these analyses requires a separation of scales between the background heterogeneity, and the emergent pattern wavelength. Additionally, there are technical obstructions to extending the ideas of \cite{krause_WKB} beyond a 1D spatial domain, though we conjecture that such a localisation result should hold on higher-dimensional spatial domains.

\begin{figure}
    \centering
    \subfloat[]{{\includegraphics[width=0.4\textwidth]{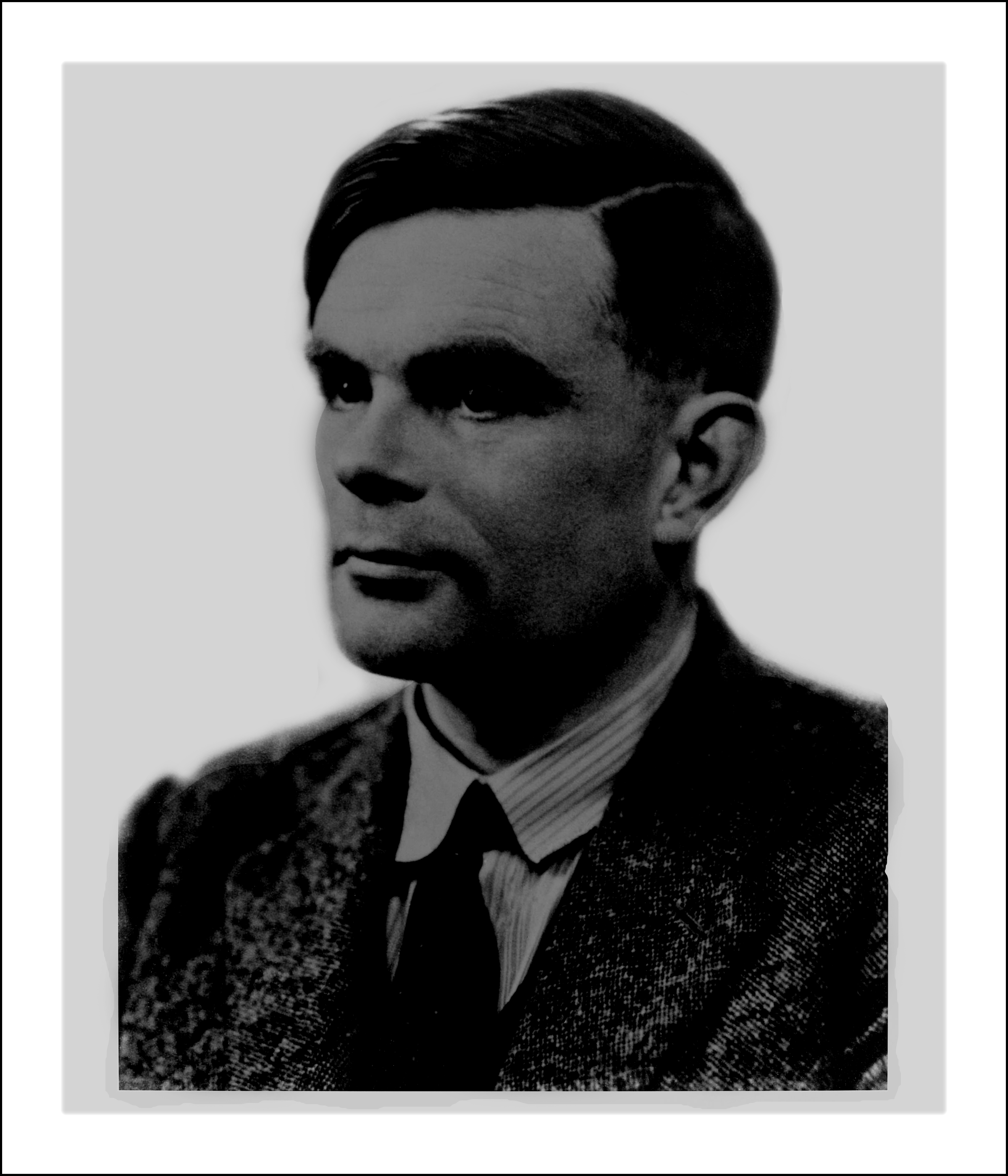}}}\hspace{0.8cm}
    \subfloat[]{\includegraphics[clip, trim = 0 0 0 4.9cm, width=0.4\textwidth]{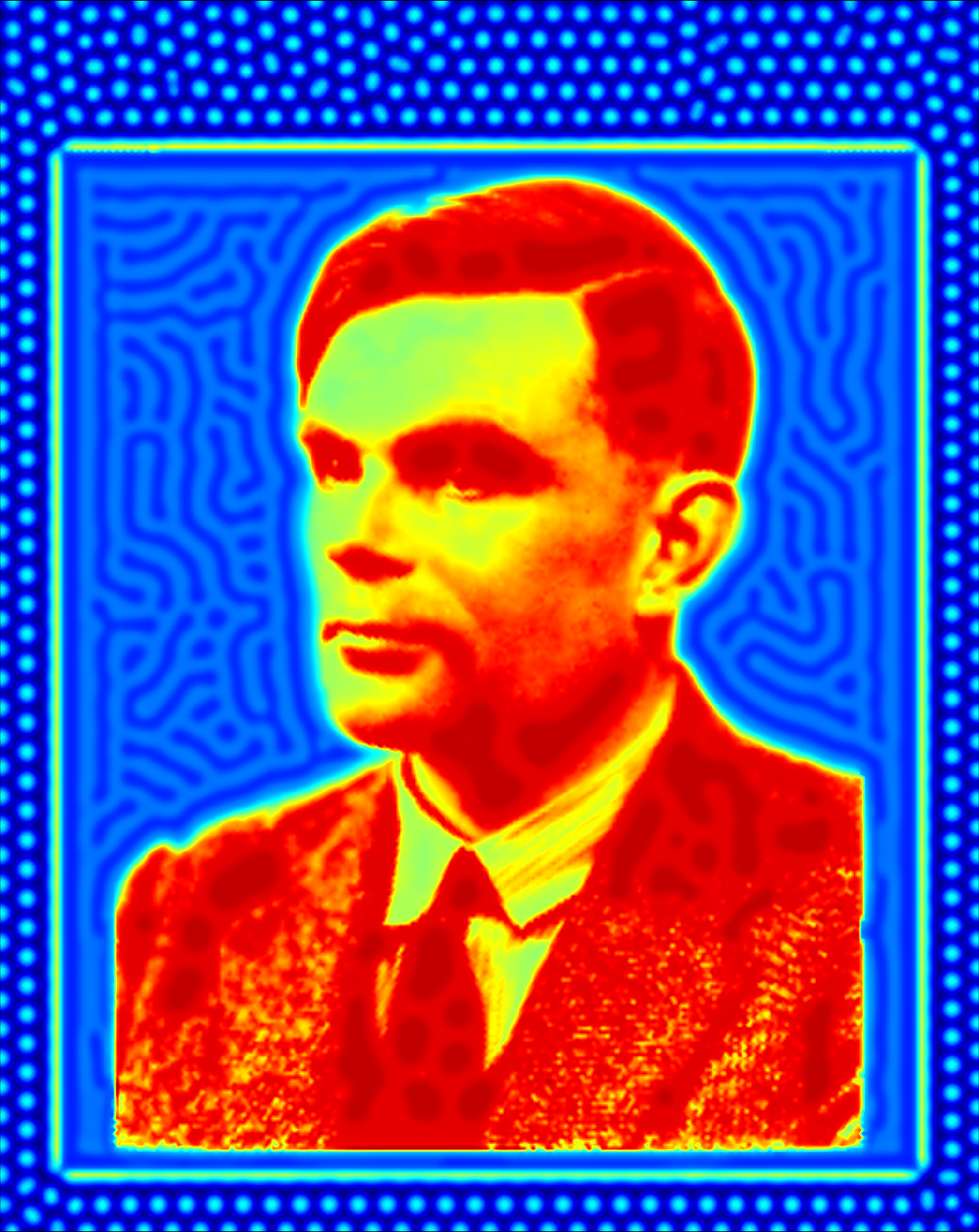}}
    \caption{(a) A black and white image of Alan Turing with three distinct regions. The outer background is white, an inner background boundary is light grey, and a central complex inner image is shades of dark grey. (b) Values of an activator concentration from simulations of a reaction-diffusion system using (a) to define a spatial heterogeneity. Figure created following the algorithm described in Figure 11 of \cite{woolley2021bespoke} } 
    \label{FigJim}
\end{figure}

This `local linear theory' has also been used to design reaction-diffusion systems which pattern in specified domains. Employing those results, as well as the stripe and spot selection criteria from \cite{ermentrout1991}, Woolley \etal \cite{woolley2021bespoke} created a framework to design reaction-diffusion systems to match a variety of patterning specifications. As an example, in Figure \ref{FigJim}, we show an image with three regions satisfying local conditions, and a simulation that uses this image as heterogeneous spatial forcing in a reaction-diffusion system. The heterogeneity in the light grey and white regions locally matches spatially homogeneous Turing systems that give rise to stripes and spots, respectively. In the innermost region, a heterogeneous base state value is stable to perturbations, and hence the concentration profile follows the complex prepattern closely. 

This `local' approach is striking in its ability to capture quite complex patterns, though we note that it is fundamentally phenomenological, and can only be justified in regimes where there is a clear separation of spatial scales. While these results are useful in showing that in such circumstances our intuition from the classical case applies, there are also examples where this intuition leads one astray. In particular, the presence of even simple spatial heterogeneity can induce spatiotemporal behaviour, such as changing the stability of patterned states and thereby inducing periodic movement of spike solutions \cite{page2005complex, krause2018heterogeneity, kolokolnikov2018pattern}. Such behaviour arises from the instability of stable spike solutions, which is beyond the scope of near-equilibrium analysis (though we note that linear stability analysis of spike solutions was used in \cite{kolokolnikov2018pattern} to understand this behaviour). We anticipate that there is much work to be done in reconciling the simple local viewpoint with these more dynamic phenomena, particularly in terms of thinking of the robustness of stationary Turing patterns in heterogeneous environments. One promising avenue is to more systematically study where separation of spatial and temporal scales can be exploited; see also \cite{Marciniak-Czochra} for more discussion on this point. Finally, we note that beyond cases of clear spatial separation, disentangling a patterned state due to diffusion and nonlinearity from one generated by environmental heterogeneity can be mathematically and philosophically difficult or intractable. This has potential implications in many areas, such as understanding when spatial clustering of animals is due to dispersal and species interactions, or environmental heterogeneity \cite{kurowski2017two, taylor_patterns_2019}. 

\subsection{Evolving Domains \& Non-Autonomous Forcing}\label{growth_sect}

As noted above, Turing's work on phyllotaxis anticipated that many pattern forming systems develop not on a static domain, but instead on one which is growing, or more generally deforming. Such systems have been studied intensively over the past 25 years or so, and led to several novel insights into Turing-type pattern formation. In some sense, the Turing instability can also be viewed as driven by growth, as it cannot occur for domains which are sufficiently small \cite{klika2018domain}. This perspective also helps clarify how a biological system can begin in a homogeneous state, and, as it grows, a pattern \emph{emerges} from a Turing instability once a certain critical domain size is exceeded. Often this first instability can be thought of as polarization, as discussed in Section \ref{structure}\ref{multidomain}. In all of the work involving growing domains discussed in this Section, we will concentrate on cases where the domain evolution is prescribed \emph{a priori}, and hence there is no feedback from the reaction-diffusion dynamics on this evolution (see \cite{lee2011dynamics} for a numerical example of concentration-dependent growth).

\begin{figure}
    \centering
    \includegraphics[width=0.6\textwidth]{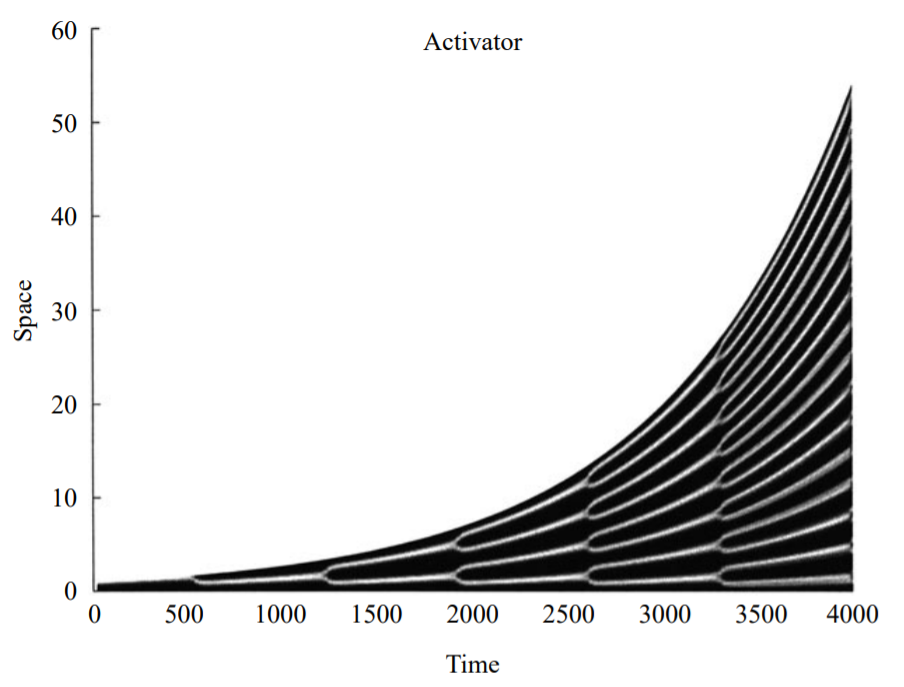}
    \caption{A period-doubling cascade in an exponentially growing reaction-diffusion system. Figure reproduced from \cite{crampin1999reaction}; used with permission.}
    \label{FigGrowth}
\end{figure}

Several authors have used this quasi-static way of viewing the influence of growth on reaction-diffusion patterning \cite{varea1997confined, murray2004mathematical}. Crampin \etal \cite{crampin1999reaction} developed a more realistic model of the influence of domain growth on reaction-diffusion processes by rederiving the governing equations via conservation of mass. They demonstrated that growth played a promising role in improving the robustness of patterned states by reducing the dependence of the pattern on the initial perturbation \cite{barrass2006mode, baker2008partial}. Depending on the rate of growth, an exact period-doubling sequence could be observed, leading to a highly regular pattern essentially independent of the initial condition; see Figure \ref{FigGrowth}. This was later extended to non-uniform growth, which could produce a variety of pattern insertion events \cite{crampin2002pattern}.  More recently, it was shown that this frequency-doubling can depend somewhat sensitively on the kind of growth rates involved, even in a 1D domain growing uniformly \cite{ueda2012mathematical}. Chemical experiments have also been carried out on growing systems using photosensitive reactions \cite{epstein}.

In terms of near-equilibrium theory of such systems, much effort has been focused on isotropically growing domains, especially in 1D. Such growth leads to a system in the Lagrangian frame which involves a time-dependent scaling of the diffusion coefficient, and an extra (nonautonomous) term due to dilution, but otherwise the system retains the same structure of a reaction-diffusion problem. For exponential isotropic growth, the dilution term in fact becomes constant, so that homogeneous steady states on a static domain can be used to find homogeneous steady states in such a growing domain with the kinetics modified by this constant term. Under the further assumption that this growth is sufficiently slow, Madzvamuse \etal \cite{madzvamuse2010stability} showed that one can obtain small but significant corrections to the classical Turing conditions, in some sense justifying quasi-static approximations mentioned above. While these analytical results were only small modifications of the classical conditions (valid for slow growth), they are significant in showing how growth can modify a system's ability to undergo Turing-type instabilities, relaxing some of the restrictions on pattern forming systems. Hetzer and Madzvamuse \cite{hetzer2012characterization} considered a type of abstract quasi-asymptotic approach to stability in time-dependent domains. Klika and Gaffney \cite{klika2017history} relaxed some of the assumptions in the previous studies through the use of Lyapunov stability, demonstrating important aspects of history-dependence in these nonautonomous systems. It was also shown that for some special kinds of growth functions, arbitrarily high wavenumbers could become unstable in finite time, indicating a breakdown of the continuum hypothesis. 

Van Gorder \etal \cite{van_gorder_growth_2019} developed a comparison-principle to determine the time-dependent linear stability of a spatially homogeneous, but possibly evolving, base state on time-dependent manifolds. An important assumption made was that the growth was locally dilational, or in other words, uniform in orthogonal directions at any point of the manifold. The base state chosen satisfies the nonautonomous system obtained when one drops the diffusion term from the reaction-diffusion system, and hence physically corresponds to the steady state concentration profiles under the effect of dilution due to growth. The resulting linear analysis then determined whether a perturbation at any point in time would lead to growth of a spatial mode, and this could be used to deduce a set of time-dependent inequalities generalizing the classical conditions for Turing instability. These inequalities were used to create time-dependent dispersion plots which, within the restrictions of a linear theory, agreed well with numerical simulations on a variety of manifolds. Several important aspects of these results were impacted by the dynamics of the homogeneous base state, which evolved independently from the reaction-diffusion system. In particular, the base state satisfies a coupled system of nonlinear ordinary differential equations, and depending on the form of the growth function, this can admit multistability, hysteresis, and various kinds of bifurcations which were reflected in simulations of the full system, and often inhibited pattern formation when they occurred.

Beyond linear instability analysis, a number of authors have also considered weakly nonlinear analyses of growing reaction-diffusion systems \cite{comanici2008patterns}, and Swift-Hohenberg equations (a related pattern-forming model) on time-dependent domains \cite{knobloch2014stability, krechetnikov2017stability}. Many open questions still remain in understanding such problems on time-varying domains, such as those raised in \cite{knobloch2015problems}. Beyond isotropic and dilational growth, non-uniform growth has been explored numerically in 1D domains \cite{crampin2002pattern} and on manifolds \cite{krause2019influence}. These studies showed a variety of impacts that non-uniform growth can have, particularly including complicated effects of hysteresis and mode selection depending on exactly how the domain evolved. In general such growth is much more difficult to investigate via analytical methods, as even if one could sensibly define a base state, one can no longer separate spatial and temporal terms from the governing equations. 

Besides time-dependent domains, substantial work has also been carried out for reaction-diffusion systems with time-dependent forcing. Timm and Okubo \cite{timm1992diffusion} use a weakly nonlinear analysis to show that for a particular predator-prey system, time-varying diffusivities negatively impact the ability of the system to support diffusion-driven (Turing) patterns. Gourley \etal \cite{gourley1996mechanisms} found similar results for small-amplitude periodic fluctuations of the diffusion coefficients. Recently several authors have studied the Turing instability of a time-dependent base state arising from a limit cycle of the reaction system \cite{challenger2015turing, kuwamura2017diffusion}. These approaches generalize more classical work on Benjamin-Feir instabilities (sometimes termed `modulational instability') of time-periodic waves studied in the Complex Ginzburg-Landau equation \cite{benjamin1967disintegration, aranson2002world}. Following from these ideas, Van Gorder \cite{van2020turing} further developed the approach presented in \cite{van_gorder_growth_2019}, generalizing both Turing and Benjamin-Feir-type instabilities to a large class of non-autonomous reaction-diffusion systems incorporating time-dependent kinetics, diffusion coefficients, and base states. This analysis was recently applied to time-varying networks \cite{van2021theory}.  

While the approach above provides a general framework for linear instability analysis of non-autonomous reaction-diffusion systems, there remain open questions. Firstly, while there is literature on understanding far-from-equilibrium phenomena (\ie the evolution and stability of large amplitude patterns, beyond where linear analysis is valid), time dependence makes many of the traditional tools used either more difficult to use, or inapplicable altogether. There are also important problems in the use of such non-autonomous systems to model a variety of physical problems. Time-dependent models introduce additional timescales, and make solution behaviours much more sensitive to initial conditions and other details which are often neglected in autonomous systems. Finally, all of the analytical work to date on growing domains has only considered the case of prescribed growth, and moving beyond prescribed growth has only been considered in numerical studies \cite{lee2011dynamics, vanderlei2011computational, buttenschon2020cell, liu2021spots, rens2021cellular}.  Many developmental systems will involve feedback between morphogen signalling and growth \cite{schwank2010regulation}, and we are unaware of any attempts to analytically study such interactions in the reaction-diffusion literature. We note in particular how difficult it is to capture concentration-dependent growth in terms of modelling assumptions, before even considering linear instability analysis etc. Nonetheless, these issues have been studied in mechanical models where domains are intrinsically coupled to the dynamics under study \cite{goriely2017mathematics}.

\section{More General Systems}\label{general}

Here we briefly sketch a few other generalizations of classical reaction-diffusion theory, namely generalized transport and models of three or more species.

\subsection{Generalized Transport Models}\label{general_transport}

The framework given by Equation \eqref{RTS} admits a few direct examples which go beyond Fickian diffusion. One example would be equations involving higher-order spatial operators, such as the Swift-Hohenberg equation \cite{swift1977hydrodynamic, cross1993pattern}, and the Cahn-Hilliard equation \cite{cahn1958free,kielhofer1997pattern}. Such higher-order operators often capture nonlocal interactions, and arise in several biological applications \cite{cohen1981generalized, ochoa1984generalized} (also see  \cite[Chapter 11]{murray2007mathematical}). These models can exhibit much more complicated dispersion relations, allowing for pattern formation in scalar systems (where one can often use variational approaches and other powerful tools to understand global aspects of the patterning process). 

Systems of reaction-advection-diffusion equations are another example of generalized transport, accounting for advection of morphogens by a variety of physical and biological mechanisms. Such systems exhibit a variety of new phenomena not present in two-component reaction-diffusion systems, such as patterns which are advected by the flow \cite{van2019diffusive}, modification of phases between activator and inhibitor \cite{satnoianu2002general}, and pattern formation due to differential-flow without the need for differential diffusion \cite{rovinsky1992chemical}. Such systems can admit both stationary and travelling waves \cite{satnoianu2003coexistence}. Reaction-advection-diffusion systems are examples of Equation \eqref{RTS} only in the case when the ratio of advection to diffusion is the same between each species, as otherwise the system can not be written or analyzed via a scalar operator $\mathcal{L}$. Even when this is possible, and the system can be studied via the spectrum of $\mathcal{L}$, the spectrum will typically be more complicated due to the loss of self-adjointness of $\mathcal{L}$, which allows complex eigenvalues and a loss of spectral discreteness in general (depending on the boundary conditions used) \cite{klika2018domain}. Reaction-advection-diffusion systems have been studied outside of this regime (that is, when the systems cannot be written using a scalar operator $\mathcal{L}$), though there is limited analysis possible from the perspective of linear instability. This is due to the fact that the theory of matrix operators, particularly when they are not self-adjoint, is not nearly as well-developed as that of the scalar Sturm-Liouville theory; see Section 3 of \cite{van2019diffusive} for further discussion of these issues.

Other transport models beyond the form of Equation \eqref{RTS} include those with nonlinear diffusion \cite{gambino2013turing, gambino2014turing}, chemotaxis and related directed-motion models \cite{horstmann20031970, maini2006turing,hillen2009user, taylor_patterns_2019},\cite[Chapter 5]{murray2004mathematical}, and cross-diffusion systems \cite{shigesada1979spatial, gambino2012turing,fanelli2013turing, madzvamuse2015cross}. We also mention hyperbolic extensions of reaction-diffusion systems, developed to account for the finite speed of propagation of real particles \cite{mendez2010reaction, zemskov2016diffusive}, though we remark that Turing-type instabilities in such systems typically exhibit nearly identical patterns in the same parameter regimes as their purely parabolic counterparts. However, these hyperbolic systems also admit oscillatory instabilities (sometimes called Turing-Wave instabilities) where a complex conjugate pair of growth rates $\lambda_k$ crosses the imaginary axis as a parameter passes the instability threshold. Such instabilities lead to spatially and temporally oscillating waves, and are not possible in two-species reaction-diffusion systems.

The framework of non-equilibrium thermodynamics offers a unifying point of view on several of these pattern forming models, where Fickian diffusion, nonlinear diffusion, Stefan-Maxwell diffusion, chemotaxis
or cross-diffusion are all particular manifestations of the same
mechanism -- the gradient of the chemical potential being a driving
force of transport processes
\cite{de2013non,krishna1997maxwell,klika2013coupling,klika2018beyond}.
From this point of view, these models differ only in the degree of coupling among such forces, analogous to the prescription of constitutive relationships in continuum mechanics. While many of these extensions to the classical Fickian model of diffusion can be derived using formalisms of nonequilibrium thermodynamics, there are important open questions regarding the appropriate model for many transport phenomena involved in pattern formation \cite{falasco2018information, esposito2020open}.

As in reaction-diffusion models, these systems exhibit diffusion-driven instabilities, as well as a range of other spatiotemporal dynamics. The usual approach for analyzing such systems near a homogeneous steady state employs an ansatz like \eqref{Ansatz} involving the eigenfunctions of the Laplacian, and determining solvability conditions analogous to the dispersion relation \eqref{disp_relation}. One major difference between transport models of these more general forms and standard reaction-diffusion systems is that new routes to spatiotemporal instabilities appear, which are not present in two-species reaction-diffusion systems. In particular, the Turing instability for a two-component reaction-diffusion system will always undergo a purely spatial instability (\ie the largest growth rate, $\lambda_k$, will be real), whereas many models involving nonlinear diffusion admit complex instability boundaries in their parameter spaces. This allows for a wider class of symmetry-breaking phenomena, and in particular Turing-wave type instabilities where spatial modes oscillate in time as their amplitudes increase.  Weakly nonlinear analysis of these more general systems is also more difficult, though progress has been made in understanding a fairly wide class of systems; see  \cite{wheeler2021convective} and the references therein.

\subsection{Beyond Two-Species}\label{beyond}
The vast majority of the work cited above considers the case of two-components, which is the simplest system that exhibits diffusion-driven instability in the class of reaction-diffusion systems. Although the linear stability analysis described in Section \ref{linear}\ref{instab_RD} applies to $m$-component systems, determining parameter ranges corresponding to instability quickly increases in complexity with the number of components, and there are new possibilities for more complicated behaviours (such as Turing-Wave bifurcations for $m \geq 3$). Some results regarding these general systems were derived by Satnoianu \etal \cite{satnoianu2000turing}. Of particular importance is the requirement that some subsystem be unstable in the absence of diffusion (\ie an activator subsystem), generalising the notion of activator-inhibitor interactions. A variety of recent approaches have analyzed these $m$-component systems from the perspective of reaction networks, in order to determine motifs that permit pattern formation \cite{marcon2016high, diego2018key}, and to explore questions of robustness in larger networks \cite{scholes2019comprehensive}. Given that the general theoretical problem is difficult to analyze, it is important to work with experimentalists on specific systems \cite{glover2017hierarchical,painter2012towards, woolley2021bespoke}.

There are important caveats linked to the study of $n$-species systems in developmental biology. Economu, Monk and Green \cite{economou2020perturbation} point to a key observation that large portions of the genome and proteome are devoted to regulation. Hence, the minimal working examples of self-organisation should be expanded as they most likely do not capture the complexity of larger systems. For instance, see \cite{klika2012influence} where a given (two-species) network can exhibit strikingly different behaviour than the same network only expanded by a single species. There are cases where more species can be valuable or unavoidable, such as in reducing the need for a substantial differential diffusion between species required for pattern formation \cite{haas2020turing}. Given that the notion of a \emph{morphogen} can be difficult to operationalize (\ie ascribe to a specific class of functional structures), there are crucial questions regarding what specifically is meant by this term in a given situation. It has been suggested in the developmental context that a reasonable level of description is the use of a specific signalling pathway as a morphogen, rather than including the complexity of a more detailed level \cite{economou2020perturbation,glover2017hierarchical}. This is an ongoing debate, and we anticipate a substantial opportunity here for mathematical ideas to help elucidate the right level of abstraction to employ in understanding developmental systems.

Another important contemporary development, also related to circumventing the need for large differential diffusion, is the study of models involving non-diffusible components within a reaction-diffusion system \cite{rauch2004role, klika2012influence, marciniak2017instability}. Such non-diffusible species often represent membrane-bound proteins or other immobile substrates, and allow for pattern formation even in the case of equal diffusion coefficients among the diffusible species \cite{lengyel1992chemical}. In some sense, this is still a kind of differential transport, and can be shown in some cases to be exactly captured by a reaction-diffusion system with different diffusion coefficients \cite{korvasova2015investigating}. Importantly, there are also requirements that the non-diffusible subsystem be stable when uncoupled from the diffusible species, in order to prevent the excitation of arbitrarily high wavenumbers, and hence a breakdown of the continuum approximation \cite{klika2012influence,korvasova2015investigating}. 

Multistability of patterns, even in simple two-component reaction-diffusion systems, can be common, especially in higher spatial dimensions \cite{borckmans1995turing}. Such complexity results purely from spatial interactions, even in systems which are monostable in the absence of diffusion. While there is important work being done in chemical reaction network theory on monostable systems (of particular note, we refer to the global attractor conjecture \cite{anderson2011proof}) it is not at all clear that biochemical signalling networks in development will have this property of monostability. The interaction of Turing-type bifurcation branches with multiple homogeneous equilibria is under-explored, and there are more avenues for such multistability to exist generically when considering larger systems.

\section{Outlook}\label{discussion}
The review presented here covers a range of different models building on Turing's ideas of pattern formation in chemical and biological systems. Much of this research directly extends ideas that were nascent in Turing's analysis, and still there are many questions implicit in Turing's analysis which remain unanswered. While Turing-type pattern formation is now well-established with a vast literature, it is also rich in biological and mathematical discoveries yet to be made. We discussed a number of advances related to geometry, heterogeneity, and reaction-transport systems above, closing each Subsection with a list of key challenges yet to be solved. However, there are further directions for exploration we have not yet discussed. We end this review highlighting a few of these. 

A difficulty alluded to above is both the identification of morphogens, and the specification of morphogen kinetics. In chemical models of Turing systems, we often have a strong indication that the chemical kinetics can be captured by specific nonlinear functions \cite{epstein}. However, such certainty is much harder to obtain in biological systems, especially due to the difficulty in operationalizing the abstract notion of \emph{morphogen}. Even in the two-component case, reaction kinetics can be designed to match a wide range of observed patterns \cite{woolley2021bespoke}. Of course, such phenomenology gives little indication that the underlying mechanism is that of a reaction-diffusion process, and there is important work to be done in carefully testing different competing modelling hypotheses in pattern formation.

There are also important uncertainties in comparing predictions from chemical theories from those from other kinds of models, such as mechanical models of pattern formation. Many of these frameworks have similar analysis, and give rise to similar principles, such as short-range activation and long-range inhibition \cite{meinhardt2000pattern}. One important aspect of differentiating between these competing theories is to specify precisely what their assumptions and predictions are, and compare these with experiments; see \cite{Marciniak-Czochra} in this Theme Issue for a discussion along these lines. Another way of rethinking these issues is to focus on conceptually separating problems in developmental biology (\eg how we explain an observation) from their solutions in terms of models or particular mechanistic hypotheses \cite{sharpe2019wolpert}. One can then see which kinds of developmental phenomena are explained by different mechanistic hypotheses, and use this to determine further experimental work which can help distinguish between competing hypotheses.

For brevity we have omitted many topics related to pattern formation and reaction-diffusion systems, but we will now briefly mention two of these. Reaction-diffusion models of the form \eqref{first_RD} are local in time, implying that reaction rates instantaneously change in proportion to the concentrations. However, these morphogens are often, though not always, signalling molecules which must be produced via gene transcription, leading to delays between uptake of a signal and the up or down regulation of the resultant product. Such time delays have been studied in the past few decades \cite{gaffney2006gene,lee2011dynamics, gaffney2015sensitivity, yi2017bifurcation, fadai2017delayed, jiang2019turing}, broadly with the conclusion that they may lead to oscillatory behaviours, rather than the formation of stationary patterns (though see \cite{fadai2018time} for a case where delay can enhance the stability of spikes). Such results plausibly call into question the standard use of reaction-diffusion frameworks for the fastest cellular self-organisation processes, where oscillations are not typically observed, though we caution that there is more work to be done both biologically and mathematically in testing these hypotheses. On the mathematical side, all of the studies to date have only considered simple models of fixed time delay, though in reality one should anticipate some distribution of time delays in noisy cellular environments. 

Another implicit simplification in the modelling frameworks explored here is to consider only mean-field deterministic models of what are typically very noisy systems. Stochastic modelling in mathematical biology, and reaction-diffusion systems in particular, has seen substantial development in the past few decades \cite{erban2019stochastic}. Such models have also been applied to study questions regarding pattern formation, both for intrinsic noise due to small number fluctuations (analytically and via Gillespie-type algorithms) \cite{woolley2011stochastic, woolley2012effects, schumacher2013noise}, and for external noise, via random forcing at the level of partial differential equations \cite{adamer2020coloured}. While noise has been shown to be both stabilizing and destabilizing in different contexts, it can play an important role in helping to address some aspects of the robustness problems alluded to earlier in the deterministic theory \cite{maini2012turing}. In particular, stochastically forced systems can explore the space of spatially-patterned states, and in some cases more easily settle in to a specific spatial configuration with less dependence on their starting conditions compared to deterministic models. Still, there is important work to be done in determining the effect of such noise, especially in the more general settings discussed in this review.

Besides these areas, we have omitted many other topics of relevance to pattern formation in reaction-diffusion systems. We have not discussed wave phenomena, which are in some sense dual to pattern formation \cite{grindrod1991patterns}, though such phenomena can also effect patterning, for example by wave pinning \cite{mori2008wave}.. Except in passing, we have also not discussed some of the rich applications of the framework of reaction-diffusion equations in spatial ecology, and the fruitful interaction these fields have had \cite{cantrell2004spatial}. Lastly, we only considered models that were continuous in space and time, though there are a range of discrete approaches, which are particularly well-motivated in considering biological cells \cite{deutsch2005cellular}. Still, we hope that this review gives the reader a sense of how far we have come from Turing's 1952 paper, and how much more work there is to be done.

Turing wrote his paper in ``the present state of knowledge," almost seventy years ago. As evidenced here, his ideas have gone on to change this state of knowledge, which has progressed iteratively both mathematically and biologically. We feel it is important to reassess this state, both to appreciate what has been accomplished, and to plan paths forward in scientific work. As Turing said (though in a  different context) \cite{machinery1950computing}, ``We can only see a short distance ahead, but we can see plenty there that needs to be done." We hope this paper has helped guide the reader to some of these near-term prospects for progress in further developing these ideas.

\enlargethispage{20pt}


\competing{The author(s) declare that they have no competing interests.}

\funding{V.K. is grateful for support from the European Regional Development Fund-Project `Center for Advanced Applied Science' (No. CZ.02.1.01/0.0/0.0/16\_019/0000778).}




\bibliographystyle{abbrv}
\bibliography{refs}

\end{document}